\documentclass[aps,preprint,prd,nofootinbib]{revtex4}

\usepackage{graphicx}
\usepackage{psfrag}

\begin{document}

\title{Production of a Heavy Quarkonium with a Photon or via ISR at $Z$ Peak in $e^+e^-$ Collider}
\author{CHANG Chao-Hsi$^{1,2,3}$\footnote{email:zhangzx@itp.ac.cn}, WANG Jian-Xiong$^{4}$\footnote{email:
jxwang@mail.ihep.ac.cn} and WU Xing-Gang$^{1}$
\footnote{email:wuxg@cqu.edu.cn}}
\address{$^1$Department of Physics, Chongqing University, Chongqing 400044,
P.R. China\\
$^2$ CCAST (World Laboratory), P.O.Box 8730, Beijing 100190, P.R. China.\\
$^3$ Institute of Theoretical Physics, Chinese Academy of Sciences, P.O.Box 2735,
Beijing 100190, P.R. China.\\
$^4$ Institute of High Energy Physics, Chinese Academy of Sciences,
P.O.Box 918(4), Beijing 100049, P.R. China }

\begin{abstract}
Considering the possibility to build an $e^+e^-$ collider at the
energies around $Z$-boson resonance with a luminosity so high as
${\cal L} \propto 10^{34}cm^{-2}s^{-1}$ (even higher) and the
abilities of a modern synthesis detector, we systematically
calculate the exclusive two body processes: $e^+e^-$ annihilates
into a heavy quarkonium and a photon (initiate state radiation i.e.
ISR is involved), at the energies around the $Z$-boson resonance.
Since the couplings of $Z$-boson to quarks contain axial vector as
well as vector, so here the produced heavy quarkonium may stand for
a charmonium such as ${J/\psi,\eta_c,h_c,\chi_{cJ}}\cdots$ and a
bottomonium such as ${\Upsilon,\eta_b,h_b,\chi_{bJ}}\cdots$
respectively. If we call such a collider with so high luminosity and
running around the $Z$-boson resonance as a $Z$-factory, then our
results obtained here indicate that experimental studies at a
$Z$-factory about the various heavy quarkona (their ground and
excited states) via the two-body processes, especially, the
production of the bound states with quantum number $J^{PC}=1^{--}$
via ISR, have outstanding advantages. \\



\end{abstract}

\maketitle

The production of a heavy quarkonium, such as a charmonium or a
bottomonium, at high energies not only may be used to test the
high-energy behavior of quantum chromodynamics (QCD) as well as the
interplay of perturbative and non-perturbative phenomena in QCD, but
also may be used to search for suitable sources to produce
experimental example for studying the properties of the heavy
quarkonia. Comparing with the hadronic colliders such as TEVATRON
and LHC, an $e^+e^-$ collider has many advantages except the low
rate of the production in studying the production and/or in using
the produced heavy quarkonium as an experimental study sample.
Recently the technique progress on $e^+e^-$ colliders motivated by
ILC has achieved, that the luminosity may be raised so high as
${\cal L}\propto 10^{34}cm^{-2}s^{-1}$ (even higher). Moreover, if
such an $e^+e^-$ collider further runs at $Z$-boson peak energy, the
resonance effects at the peak may raise the production rate to
several magnitude order high, thus the shortcoming (the low
production rate) can be overcome much for the mentioned purposes.
For convenience, we will call as a collider as a `$Z$-factory' later
on\footnote{In the literature (Ref.\cite{gigaz} and references
therein) running the prospective high-energy $e^+ e^-$ collider is
called as GigaZ.}. As estimated in Ref.\cite{gigaz}, in terms of
GigaZ, one can perform experiments on the basis of an example more
than $10^{9\sim 10}$ $Z$-events, that enlarges the number of
$Z$-events by $3\sim 4$ orders of magnitude than that collected in
LEP-I. Thus we think $Z$-factory will be able to open new
opportunities not only for high precision physics in the
electro-weak sector, but also for the hadron physics including that
of the heavy quarkonia.

Considering the abilities of a modern synthesis detector for
photons, and the fact that the energy-momentum is monochromatic in a
two-body process, who's final state contains two body only and
incoming energy is fixed, in the present paper we would like to
focus the two body processes for the heavy quarkonium production as
below:
\begin{equation}
e^+(p_1) + e^-(p_2) \rightarrow {\gamma}(p_3) + H_{Q\bar{Q}}(P)
\end{equation}
where the heavy quarkonium state $H_{Q\bar{Q}}$ stands for
${J/\psi}$, ${\eta_c}$, ${h_c}$, ${\chi_{cJ}}$ ($J=0,1,2$) for
charmonium and stands for ${\Upsilon}$, ${\eta_b}$, ${h_b}$,
${\chi_{bJ}}$ ($J=0,1,2$) for bottomonium respectively\footnote{In
fact, here $H_{Q\bar{Q}}$ can stand for higher excited states too,
but in the present paper we restrict ourselves to consider the
low-lying excited states only.}, especially at the energy of
$Z$-peak. Since the produced heavy quarkonium in the two-body
processes is monochromatic in energy-momentum, so experimentally the
heavy quarkonium may be comparatively easy to be identified, as long
as the investigated event-sample is big enough. Thus how great the
cross sections are crucial for experimental studies of the heavy
quarkonia. Moreover to observe the production of the heavy quarkonia
at $Z$-peak via the two body process, the resonance effects i.e. the
contributions from `$Z$-boson exchange' become greater than those
from `virtual-photon exchange' owing to the $Z$-boson propagator
comes to its mass-shell, and `$Z$-boson exchange' is of weak
interaction (without parity and charge-parity conservation), hence
various states of quarkonia with different parity and charge-parity
can be produced quite equally. To the leading order calculation,
there are four typical Feynman diagrams FIG.1a, FIG.1b, FIG.1c and
FIG.1d for the processes (since each typical Feynman diagram in
FIG.1 contains two diagrams: a virtual-photon exchange one and a
virtual $Z$-boson exchange one, so the total number of the
independent Feynman diagrams represented in FIG.1 is eight in fact)
if the the heavy quarkonia in the final state have quantum numbers
as $J^{PC}=1^{--}$, whereas there are two FIG.1c and FIG.1d only
(for the same reason as in the case for the heavy quarkonia
$J^{PC}=1^{--}$, the number of independent Feynman diagrams are
four) if the the heavy quarkonia in the final state have quantum
numbers different from $J^{PC}=1^{--}$. As matter of fact, FIG.1a
and FIG.1b are those of ISR (here one may see that ISR produce the
states with quantum numbers $J^{PC}=1^{--}$ only). Note here that
due to the fact that a heavy-quark propagator via two vertices
connects with the heavy quarkonium as shown in the diagrams FIG.1c
and FIG.1d, but not as that shown in the diagrams FIG.1a and FIG.1b,
a photon directly connects to the heavy quarkonium via a single
vertex, then the heavy quarkonium, involving orbital angle-momentum
excitation ($L>0$), can be also produced via this mechanism depicted
by the diagrams FIG.1c and FIG.1d.

\begin{figure}
\includegraphics*[scale=0.5]{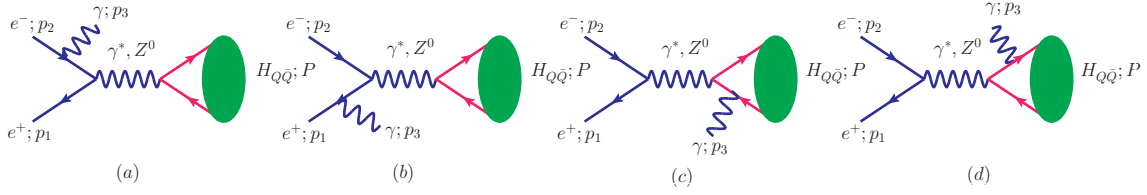}
\caption{Typical Feynman diagrams for the process $e^+(p_1) +
e^-(p_2) \rightarrow \gamma^*/Z^0 \rightarrow {\gamma}(p_3) +
H_{Q\bar{Q}}(P)$, where $H_{Q\bar{Q}}$ equals to ${J/\psi}$,
${\eta_c}$, ${\chi_{cJ}}$ ($J=0,1,2$) for charmonium and
${\Upsilon}$, ${\eta_b}$, ${\chi_{bJ}}$ ($J=0,1,2$) for bottomonium
respectively. }\label{feynman}
\end{figure}

The calculations on the processes, no matter for the ground state or
for the excited states, need to establish the exact relations
between each term of the amplitude (each Feynman diagram) and the
nonperturbative matrix elements in NRQCD framework\cite{nrqcd}, or
equivalently, the relevan Bethe-Salpeter (B.S.) wave function
(derivative of the wave function) at origin for the $S$-($P$-)wave
production respectively in B.S framework\cite{mand}.

Generally, according to Feynman rules, the term of the amplitude
relating to the concerned Feynman diagram, can be written as
\begin{equation}\label{amp}
M_{k}={\cal C}\bar{v}_{e}(p_1)\Gamma_{1k} u_{e}(p_2)
i\int\frac{d^{4}q} {(2\pi)^{4}} Tr\left[\bar{\chi}(P,q)
\Gamma_{2k}\right] \;,
\end{equation}
where $\bar{\chi}(P,q)$ is the B.S wave function for the concerned
quarkonium, ${\cal C}$ stands for an overall parameter, which is the
same for all the Feynman diagrams. $\Gamma_{1k}$ stands for the
structure of the $k$-th Feynman diagram, and is for the electric
line, which includes the relative string of Dirac $\gamma$ matrices
and the corresponding scalar part of the propagators, $\Gamma_{2k}$
stands for the similar structure of the heavy quark line. For
convenience, the intermediate photon or $Z$ propagator is put into
$\Gamma_{1k}$.

As for a quarkonium in $S$-wave state, in the non-relativistic
approximation, $ \bar{\chi}(P,q) $ can be written as
\begin{equation}
\bar{\chi}(P,q)=\phi(q)\frac{1}{2\sqrt{M}}(\alpha\gamma_5 +
\beta\hat\epsilon(S_{z}))(\hat{P}+M)\,,
\end{equation}
where $P$, $M\simeq m_{Q}+m_{\bar{Q}}$ are the momentum, the mass of
the heavy quarkonium respectively and $q$ is the relative momentum
between the two constituent quarks. In this paper, we use $\hat{a}$
to denote the contraction between the Dirac $\gamma$ matrix and a
momentum or polarization vector $a$, i.e. we use $\hat{a}$ instead
of $\slash\!\!\!a$. $\alpha=1$, $\beta=0$ for the pseudoscalar
$([^{1}S_{0}])$ quarkonium and $\alpha=0$, $\beta=1$ for the vector
$([^{3}S_{1}])$ one. The wave function $\phi(q)$, the radial part of
the momentum space, can be related to the space-time wave function
at origin by the integration, $i\int\frac{d^{4}q}{(2\pi)^{4}}
\phi(q) =\psi(0)$, where $|\psi(0)|^2=|R(0)|^2/(4\pi)$. Then the
$S$-wave amplitude can be simplified as
\begin{equation}
M_{k}={\cal C}\bar{v}_{e}(p_1)\Gamma_{1k} u_{e}(p_2) \psi(0)
Tr\left[\frac{1}{2\sqrt{M}}(\alpha\gamma_5 +
\beta\hat\epsilon(S_{z}))(\hat{P}+M) \Gamma^0_{2k}\right] \;,
\end{equation}
where $\Gamma^0_{2k}$ is defined by the following Eq.(\ref{expand}).

As for a quarkonium in $P$-wave state, in the non-relativistic
approximation, the B.S. wave function
$\bar{\chi}(P,q)=\bar{\chi}^{[^{2S+1}P_JJ_z]}(P,q)$ can be written
as
\begin{eqnarray}\label{bound}
&\displaystyle \bar{\chi}^{[^{(2S+1)}P_J,J_z]}(P,q) \simeq
\sum_{S_z,\lambda,\lambda'}\frac{-\sqrt{M}}{4m_Q m_{\bar{Q}}}\cdot
\Psi(q)\cdot(\epsilon^\lambda \cdot q)\cdot\nonumber \\
&(\hat{q}_{\bar{Q}}-m_{\bar{Q}})\cdot
(\delta_{S,0}\delta_{S_z,0}\gamma_5
+\delta_{S,1}\delta_{S_z,\lambda'}\hat\epsilon^{\lambda'})\cdot
(\hat{q}_{Q}+m_{Q})\cdot \langle 1\lambda;SS_z |JJ_z \rangle,
\end{eqnarray}
where $q_{Q}$ and $q_{\bar{Q}}$ are the momenta of $Q$ and $\bar{Q}$
quarks inside the quarkonium:
\begin{equation}
q_{\bar{Q}}=\alpha_1 P-q\,,\;\;\; q_{Q}=\alpha_2
P+q\,,\;\;\;\;\alpha_1=\frac{m_{\bar{Q}}}{m_{Q}+m_{\bar{Q}}}\,,\;\;\;
\alpha_2=\frac{m_{Q}}{m_{Q}+m_{\bar{Q}}}\,. \label{eq:momentum}
\end{equation}
For convenience, we introduce $q^{\mu}_{\perp}\equiv
q^{\mu}-q^\mu_{\parallel}$, $q^{\mu}_{\parallel}\equiv \frac{(P\cdot
q)}{M^2}P^{\mu}$ and $q_{\parallel}\equiv |q^{\mu}_{\parallel}|$,
hence accordingly we have $d^4q=dq_{\parallel}d^{3}q_{\perp}$.
$\Psi(q)$ stands for the `$P$-wave scalar wave function', and
$\langle 1\lambda; SS_z|JJ_z \rangle$ is the Clebsch-Gordon
coefficient for L-S coupling. The spin structure of the B.S. wave
function $\bar{\chi}^{[^{(2S+1)}P_J,J_z]}(q) $ defined in
Eq.(\ref{bound}) is of the lowest order in $q$) for the $P$-wave
state production.

The derivative of the wave function at origin $\psi'_{0}(0)$ in
coordinate representation relates to the $P$-wave scalar function
$\Psi(q)$ in Eq.(\ref{bound}) under the so-called instantaneous
approximation by the integration:
\begin{eqnarray}
\label{pwavezero}
i\int\frac{dq_{\parallel}d^{3}q_{\perp}}{(2\pi)^{4}}q^{\alpha}
\Psi(q)(\epsilon^\lambda \cdot
q)=i\int\frac{d^{3}q_{\perp}}{(2\pi)^{3}}\tilde{q}_{\perp}^{\alpha}
\phi\left(-\frac{q^2_{\perp}}{M^2}\right)(\epsilon^\lambda \cdot
q_{\perp})=i\epsilon^{\lambda\;\alpha} \psi^{\prime}(0)\,,
\end{eqnarray}
where $|\psi'(0)|^2=|R'(0)|^2/(4\pi)$. Here ${\tilde{q}}_{\perp}$ is
the unit vector $\frac{q_{\perp}}{\sqrt{-q^2_{\perp}}}$, and
$\phi(-\frac{q^2_{\perp}}{M^2})$ stands for the $P$-wave `scalar
wave function' in the sense of potential model. Substituting
Eq.(\ref{eq:momentum}) into Eq.(\ref{bound}), we obtain
\begin{eqnarray}\label{eq:ab}
&\displaystyle
i\int\frac{dq_{\parallel}}{(2\pi)}\bar{\chi}^{[^{(2S+1)}P_J,J_z]}(P,q)\simeq
\sum_{S_z,\lambda,\lambda'}\phi\left(-\frac{q^2_{\perp}}{M^2}\right)
(\epsilon^\lambda \cdot q_{\perp})\langle 1\lambda ;SS_z | JJ_z \rangle \nonumber \\
&\displaystyle\cdot\Big\{\frac{1}{2\sqrt{M}}(-\hat{P}+M)\cdot
(\delta_{S,0}\delta_{S_z,0}\gamma_5
+\delta_{S,1}\delta_{S_z,\lambda'}\hat\epsilon^{\lambda'})\nonumber\\
&\displaystyle-\Big(\frac{\sqrt{M}}{4m_Q m_{\bar{Q}}}\Big) \cdot
\Big[\alpha_2\hat{q_{\perp}}
(-\hat{P}+M)(\delta_{S,0}\delta_{S_z,0}\gamma_5
+\delta_{S,1}\delta_{S_z,\lambda'}\hat\epsilon^{\lambda'})\nonumber\\
&\displaystyle+\alpha_1(-\hat{P}+M)\cdot(\delta_{S,0}\delta_{S_z,0}\gamma_5
+\delta_{S,1}\delta_{S_z,\lambda'}\hat\epsilon^{\lambda'})\hat{q_{\perp}}\Big]
+{\cal O}(q_{\perp}^2)\Big\}\nonumber\\
&\displaystyle=\sum_{S_z,\lambda,\lambda'}\phi\left(-\frac{q^2_{P\perp}}{M^2}\right)
(\epsilon^\lambda \cdot q_{\perp}) \langle 1\lambda ;SS_z | JJ_z
\rangle (A^{\lambda'}_{SS_z}
+B^{\lambda'\;\mu}_{SS_z}q_{{\perp}\mu}\nonumber \\
&\displaystyle +{\cal O}(q_{\perp}^2)),
\end{eqnarray}
where
\begin{eqnarray}
&\displaystyle A^{\lambda'}_{SS_z}\equiv
\frac{1}{2\sqrt{M}}(-\hat{P}+M)
\cdot(\delta_{S,0}\delta_{\lambda',0}\gamma_5
+\delta_{S,1}\delta_{S_z,\lambda'}\hat\epsilon^{\lambda'})\,,\nonumber\\
&\displaystyle B^{\lambda'\;\mu}_{SS_z}\equiv
-\left(\frac{\sqrt{M}}{4m_Q m_{\bar{Q}}}\right) \cdot
\Big[\alpha_2\gamma^{\mu}
(-\hat{P}+M)\cdot(\delta_{S,0}\delta_{\lambda',0}\gamma_5
+\delta_{S,1}\delta_{S_z,\lambda'}\hat\epsilon^{\lambda'})
\nonumber \\
&+\alpha_1(-\hat{P}+M)\cdot(\delta_{S,0}\delta_{\lambda',0}\gamma_5
+\delta_{S,1}\delta_{S_z,\lambda'}\hat\epsilon^{\lambda'})
\gamma^{\mu}\Big]\,.\nonumber
\end{eqnarray}
To leading order of the relativistic approximation and according to
Eq.(\ref{amp}), the next step is to do the expansion of the
$\Gamma_{2k}$ about $q_\mu$ up to ${\cal O}(q^2)$ for the $P$-wave
production, i.e.
\begin{equation}\label{expand}
\Gamma_{2k}=\Gamma^0_{2k}+\Gamma^{\mu}_{2k}\cdot q_{\mu}+{\cal
O}(q^2)\,,
\end{equation}
where $\Gamma^0_{2k}$ and $\Gamma^{\mu}_{2k}$ are independent of
$q_\mu$. Substituting the above equations into Eq.(\ref{amp}) and
carrying out the integration over
$d^4q=dq_{\parallel}d^{3}q_{\perp}$ and with the help of
Eq.(\ref{pwavezero}), we obtain
\begin{eqnarray}\label{eq:m}
&M_{k}= {\cal C}\bar{v}_{e}(p_1)\Gamma_{1k}
u_{e}(p_2)\sum_{S_z,\lambda,\lambda'}\langle 1\lambda;SS_z |JJ_z
\rangle \psi'(0)\epsilon^\lambda_{\mu}\cdot
Tr\left[A^{\lambda'}_{SS_z}\cdot \Gamma^{\mu}_{2k} +
B^{\lambda'\mu}_{SS_z}\cdot \Gamma^0_{2k}\right].
\end{eqnarray}

\begin{table}
\begin{center}
\caption{Functions $C^{^{(2S+1)}P_J}$, $E_{\mu}^{[^{2S+1}P_J]J_z}$
and $F^{[^{2S+1}P_J]J_z}$ for a particular $P$-wave heavy quarkonium
states, where $\epsilon^\lambda_{\mu}$ and $\epsilon^{J_z}_{\mu\nu}$
are polarization vector and tensor respectively. }\vspace{4mm}
\begin{tabular}{||c||c|c|c||}
\hline\hline ~$P$-wave state~ & ~$E_{\mu}^{[^{2S+1}P_J]J_z}$~ &
~$F^{[^{2S+1}P_J]J_z}$~ & ~$C^{^{(2S+1)}P_J}$~ \\
\hline\hline $^1P_1$ & $(-\hat{P}+M)\gamma_5\epsilon^\lambda_{\mu}$
& $\frac{-\hat{P}\hat{\epsilon}^\lambda\gamma_5+M(\alpha_1-
\alpha_2)\hat{\epsilon}^\lambda\gamma_5}{2M\alpha_1\alpha_2}$
& $\frac{1}{2\sqrt{M}\sqrt{N_c}}$ \\
\hline $^3P_0$ & $(-\hat{P}+M)(M\gamma_{\mu}+P_{\mu})$  &
$-\frac{3[(\alpha_2-\alpha_1)\hat{P}+M]}{2\alpha_1\alpha_2}$
& $\frac{1}{2\sqrt{3N_c}M^{3/2}}$ \\
\hline $^3P_1$ & $i(\hat{P}+M)\epsilon_{\mu}^{\;\;\;\rho\varrho\nu}
\gamma_{\rho}P_{\varrho}\epsilon^\lambda_{\nu}$  &
$-\frac{\hat{P}\hat{\epsilon}^\lambda\gamma_5
-M\hat{\epsilon}^\lambda\gamma_5}{\alpha_1\alpha_2}$
& $\frac{1}{2\sqrt{2N_c}M^{3/2}}$ \\
\hline $^3P_2$ & $(-\hat{P}+M)\gamma_5\gamma_{\nu}\epsilon_{\mu\nu}^{J_z}$
& $0$ & $\frac{1}{2\sqrt{N_cM}}$  \\
\hline\hline
\end{tabular}
\label{tab}
\end{center}
\end{table}

For a specific $P$-wave state, summing over the explicit $\lambda$
and $\lambda'$ for the Clebsch-Gordon coefficients, Eq.(\ref{eq:m})
can be further simplified as
\begin{equation}\label{eq:simmatrix}
M_{k}^{S,JJ_z}={\cal C}\bar{v}_{e}(p_1)\Gamma_{1k}
u_{e}(p_2)C^{^{2S+1}P_J}\psi'(0)Tr \left[
E_{\mu}^{[^{2S+1}P_J]J_z}\cdot \Gamma^{\mu}_{2k} +
F^{[^{2S+1}P_J]J_z}\cdot \Gamma^0_{2k}\right],
\end{equation}
where the overall factor $C^{^{(2S+1)}P_J}$ and the functions
$E_{\mu}^{[^{2S+1}P_J]J_z}$ and $F^{[^{2S+1}P_J]J_z}$ are shown in
TAB.\ref{tab}.

To do the computation we adopt the `linear polarization' for the
heavy quarkonium vector and tensor states accordingly, and their
explicit formulation can be found in Ref.\cite{bcpwave}. The
amplitude of all mentioned processes can be easily generated by
using FDC\footnote{The relevant Fortran program for the mentioned
processes is available upon request.}. To simplify the amplitude,
firstly we establish a complete set of ` basic spinor lines', such
as that they are constructed by the multiplication of Dirac $\gamma$
matrixes in a certain way as $\hat p_1\hat p_2 \hat p_3 \cdots$,
then we consider them as bases to expand every terms of the
amplitude and sum up all the terms according to the expansion (the
coefficients of the `bases' are summed respectively). By this way,
the amplitude and its square become quite condense, thus the
efficiency for computing the amplitude squared is raised greatly.
This kind of treatments can be carry out quite automatically in FDC
package now\cite{fdc}.

When doing numerical calculations, we adopt
\begin{eqnarray}
&&m_{Z}=91.187GeV \;,\; m_b=4.9GeV \;,\; m_c=1.5GeV\;,
\end{eqnarray}
and
\begin{eqnarray}
&&|R^{(c\bar{c})}_{1S}(0)|^2=0.810GeV^3\;,\;
|R^{(b\bar{b})}_{1S}(0)|^2=6.447GeV^3\;,\;\nonumber\\
&& |R^{(c\bar{c})}_{2P}(0)|^2=0.075GeV^5\;,\;
|R^{(b\bar{b})}_{2P}(0)|^2=1.417GeV^5
\end{eqnarray}
for the parameters of heavy quarkonia \cite{spectrum}. In order to
have some idea about more samples to produce a `particle' with such
quantum numbers as $J^{PC}=1^{--}$ via the ISR in addition to the
ground states $J/\psi$ and $\Upsilon$,

\begin{figure}
\includegraphics[scale=0.4]{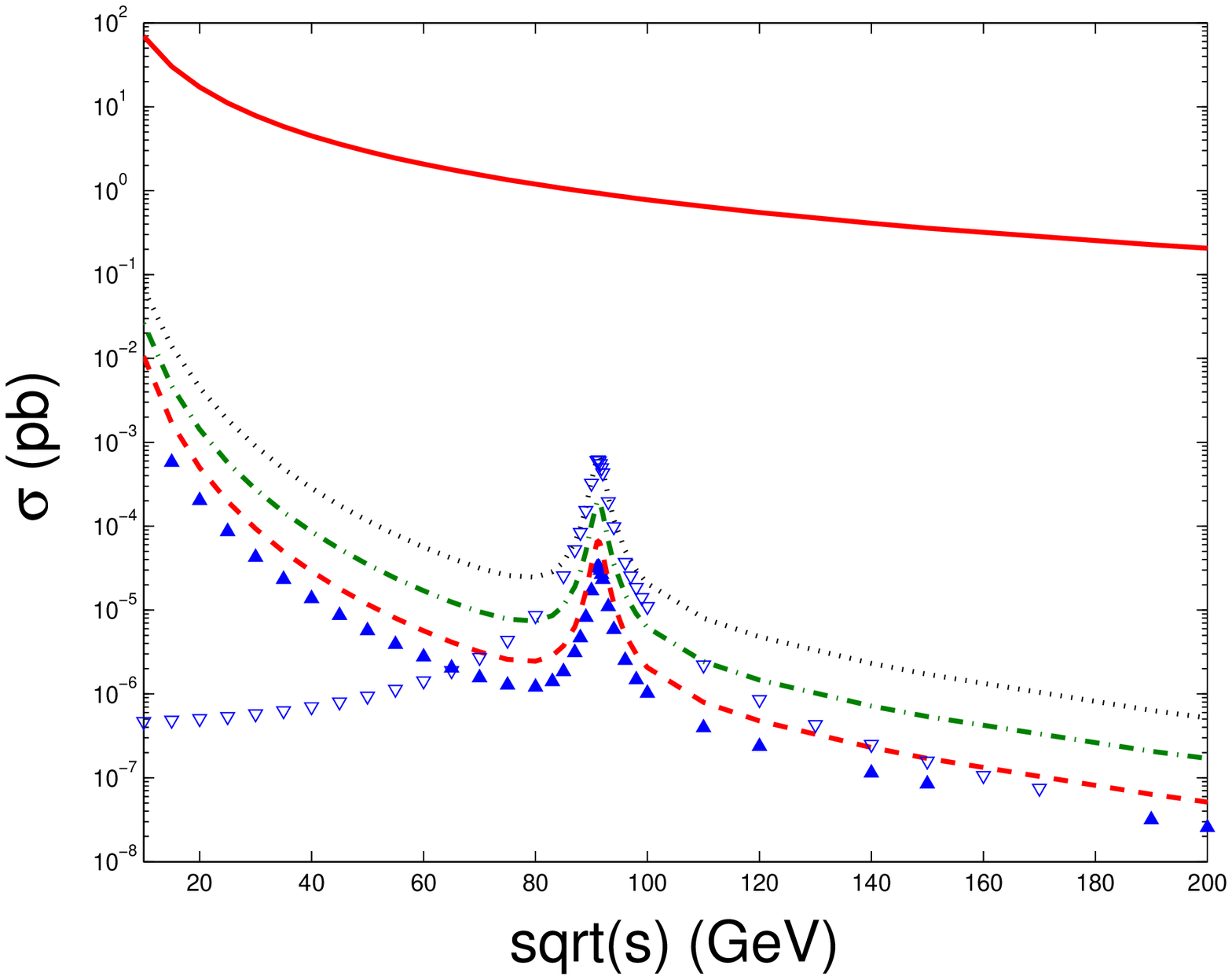}
\includegraphics[scale=0.4]{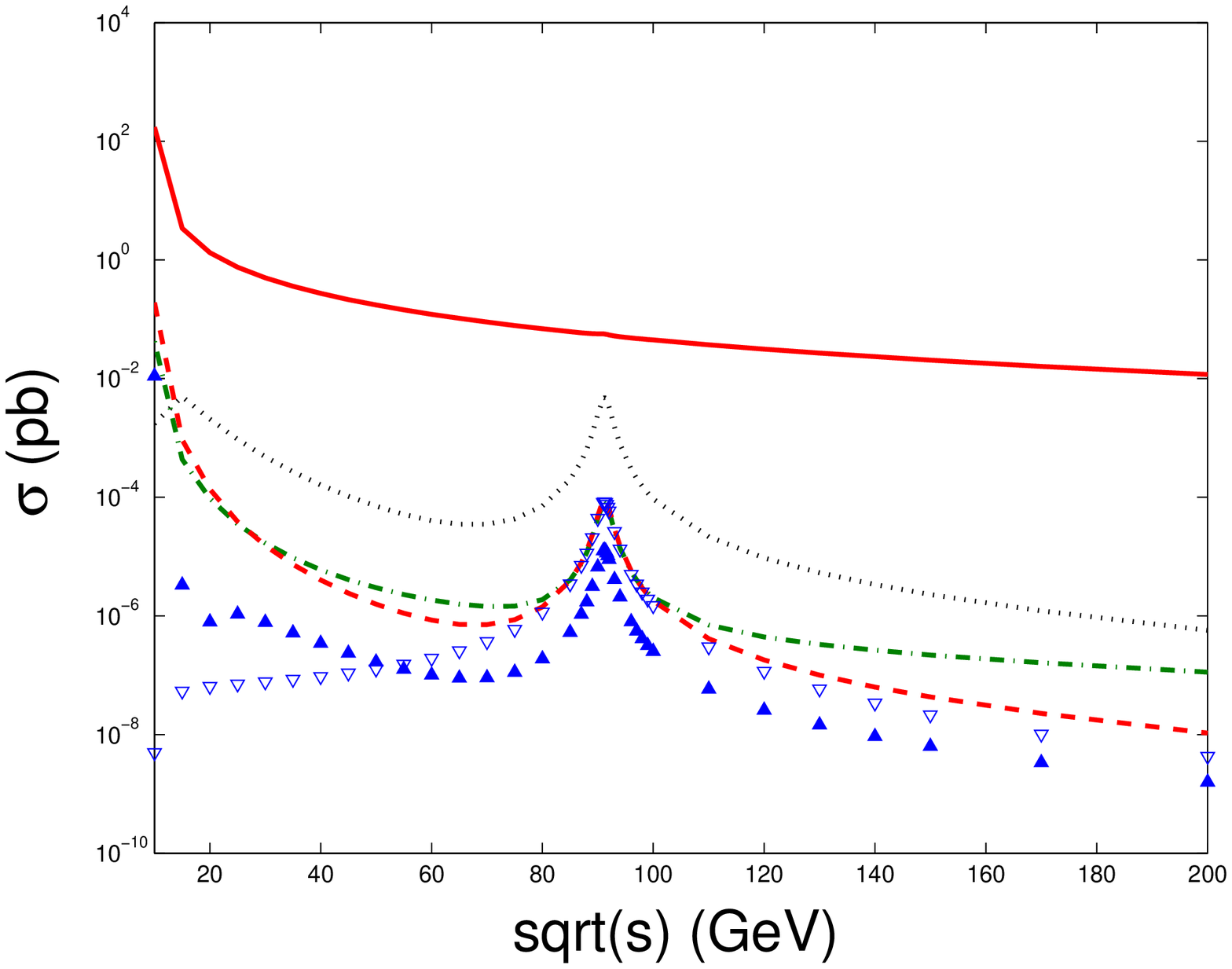}
\caption{(color online) Total cross sections for the processes $e^-
+ e^+ \rightarrow {\gamma} + H_{Q\bar{Q}} $ versus the collision
energy. The red solid, the black dotted, the blue up-solid-triangle,
the green dash-dotted, the red dashed and the down-hollow-triangle
lines stand for $Q\bar{Q}$ in $^3S_1$, $^1S_0$, $^3P_0$, $^3P_1$,
$^3P_2$, $^1P_1$ respectively. The left figure is for charmonium
and the right one is for bottomonium.} \label{totaldis}
\end{figure}

\begin{figure}
\includegraphics[scale=0.4]{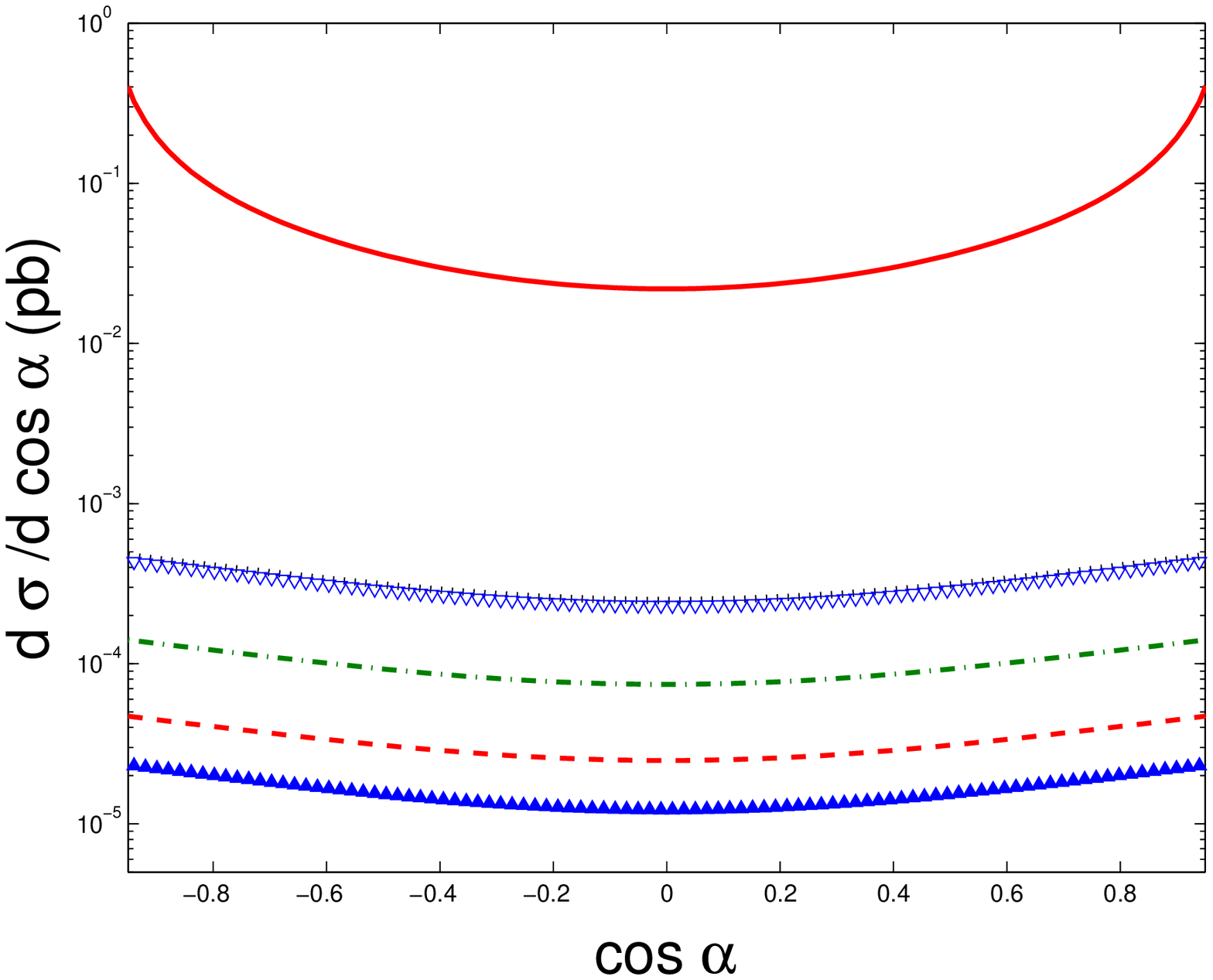}
\includegraphics[scale=0.4]{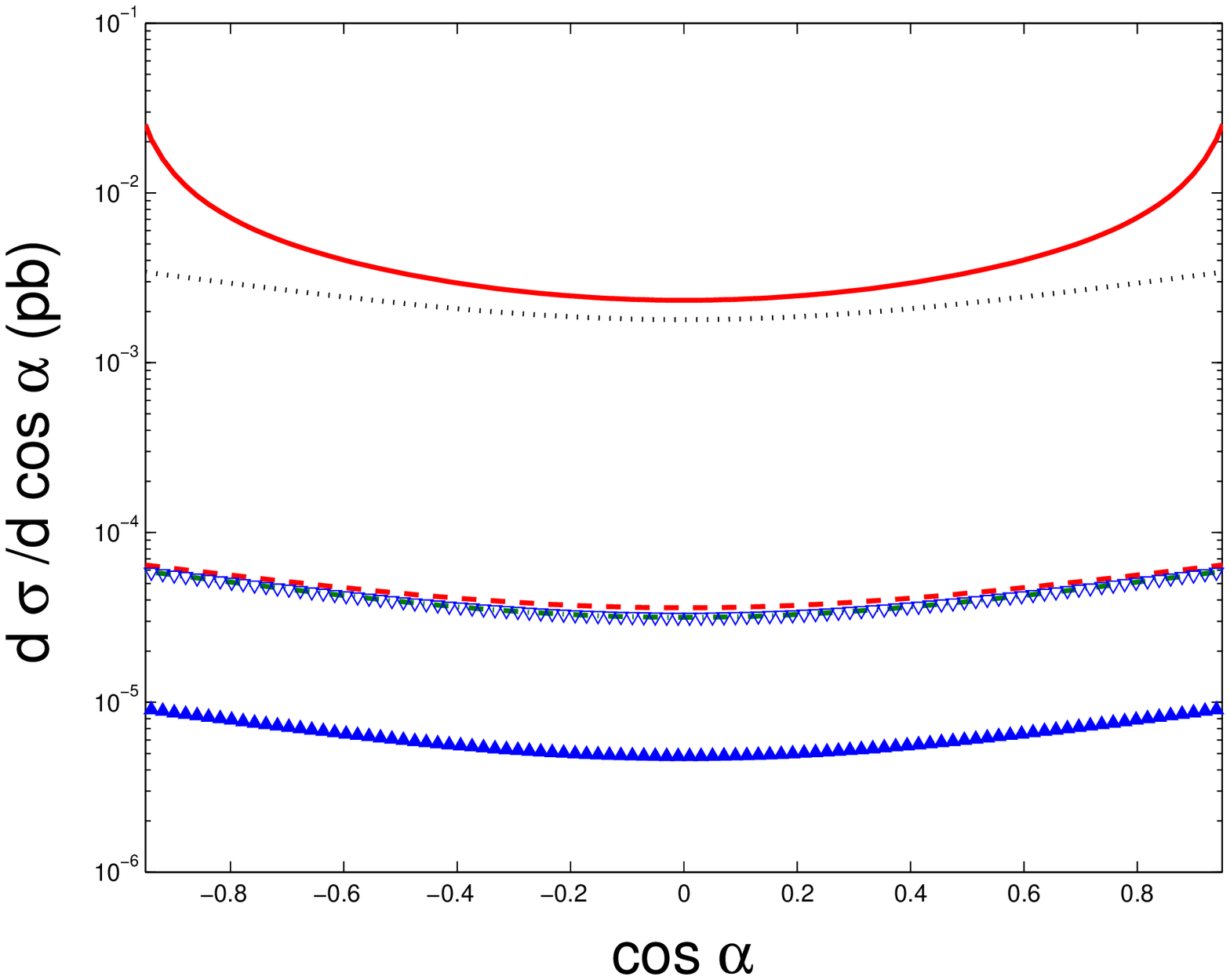}
\caption{(color online) Differential cross sections for the
processes $e^- + e^+ \rightarrow {\gamma} + H_{Q\bar{Q}} $ versus
$\cos\alpha$ at a C.M.S. energy as $Z$-mass. The red solid, the
black dotted, the blue up-solid-triangle, the green dash-dotted, the
red dashed and the blue down-hollow-triangle lines stand for
$Q\bar{Q}$ in $^3S_1$, $^1S_0$, $^3P_0$, $^3P_1$, $^3P_2$, $^1P_1$
respectively. The left figure is for charmonium (the dotted line and
the blue down-hollow-triangle almost emerge together almost) and the
right one is for bottomonium (the red dashed line, the green
dash-dotted line and the blue down-hollow-triangle emerge together
almost).} \label{cosdis}
\end{figure}

\begin{table}
\begin{center}
\caption{Total cross sections for the different quarkonium states at
the collision energy $\sqrt{s}=m_{Z}$.} \vspace{4mm}
\begin{tabular}{||c||c|c|c|c|c|c||}
\hline\hline ~~~ & ~~$^3S_1$~~ & ~~$^1S_0$~~& ~~$^3P_0$~~
& ~~$^3P_1$~~ & ~~$^3P_2$~~ & ~~$^1P_1$~~  \\
\hline\hline $\sigma_{(c\bar{c})} (pb)$ & $0.934$ &
$0.662\times10^{-3}$ & $0.328\times10^{-4}$
& $0.197\times10^{-3}$ & $0.661\times10^{-4}$ & $0.615\times10^{-3}$ \\
\hline
$\sigma_{(b\bar{b})} (pb)$ & $0.565\times10^{-1}$ & $0.475\times10^{-2}$
& $0.128\times10^{-4}$ & $0.838\times10^{-4}$ & $0.930\times10^{-4}$
& $0.833\times10^{-4}$\\
\hline\hline
\end{tabular}
\label{cross}
\end{center}
\end{table}

To present the results, the dependence of the total cross sections
for the processes $e^+ + e^- \rightarrow {\gamma} + H_{Q\bar{Q}}$
with $H_{Q\bar{Q}}$ for charmonium and bottomonium on its collision
energy is drawn in FIG.\ref{totaldis}, and the dependence of the
differential cross sections for the processes $e^- + e^+ \rightarrow
{\gamma} + H_{Q\bar{Q}}$ on $\cos\alpha$, where $\alpha$ stands for
the angle between the quarkonium and the beam, is drawn in
FIG.\ref{cosdis}. From FIG.\ref{totaldis}, one can see clearly that
as expected there is a peak at $\sqrt{s}=m_{Z}$ for the quarkonium
states of $^1S_0$, $^1P_1$ and $^3P_J$ etc. It is because that in
these processes the $s$-channel diagrams of $Z$-boson exchange
FIG.1c and FIG.1d become dominant, when the $Z$-propagator in the
diagrams approaches to mass-shell. We should emphasize that the
$Z$-boson coupling is not only of vector, but also of axial vector,
so the states with various quantum numbers, as long as they are
lighter than $m_Z$, may be produced in terms of the $s$-channel
diagrams of $Z$-boson exchange, and there must be a peak at
$\sqrt{s}=m_{Z}$. In contrary, if the production is not at $Z$-mass
energy but at an energy much low, then of the $s$-channel diagrams
FIG.1c and FIG.1d, $Z$-boson exchange ones become ignorable and only
the ones of $\gamma$ exchange play a role, so that there is a vector
coupling only, and quite a lot heavy quarkonium states without
suitable quantum numbers cannot be produced in terms of the
`$s$-channel mechanism'.

However, for a hidden flavored `comparatively light' quarkonium
($m_{H_{Q\bar{Q}}}\ll m_Z$), when the quarkonium is a $^3S_1$ or
$^3D_1$ state i.e. that with quantum numbers as $J^{PC}=1^{--}$,
then the t-channel photon-exchange diagrams, as shown in FIG.1a and
FIG.1b, will be dominant over the other diagrams (including FIG.1c
and FIG.1d even the $Z$-propagator being on-shell and the t-channel
Z-boson-exchange ones in FIG.1a and FIG.1b). In fact, the
`dominance' is also the nature of ISR essentially. Thus the
production of a $J^{PC}=1^{--}$ heavy quarkonium is very large and
without such a peak behavior around the energy of $Z$-boson as
emerged in the production of some other states. To show the fact,
the dependence of the cross-sections for producing this kind of
quarkonia is also drawn in FIG.\ref{totaldis}.

Moreover, to make comparison, we put the total cross-sections for
the production of the all low-lying states at $Z$-factory in
TAB.\ref{cross}. We would like to note here that if varying
$\sqrt{s}=m_{Z}\pm0.5$ GeV slightly, the results  will bring $\pm
1\%$ difference in the total cross sections.

From FIG.\ref{cosdis}, one may see that the distributions for the
production of the heavy quarkonia are quite flat in directions,
although in the forward and backward directions they arise in
certain aspect, thus there is not essential loss on observing the
heavy quarkonia, if putting a necessary cut around the beam
directions.

To see the role quantitatively that ISR may play in the two-body
final state processes at $Z$-factory, we assume a '$X(4260)$-like'
particle with quantum numbers $J^{PC}=1^{--}$ and calculate its
production precisely. For convenience, we denote the
'$X(4260)$-like' particle as $X_{b\bar{b}}$ below, and for
definiteness we further assume its mass is $m_{X_{b\bar{b}}}\simeq
15$ GeV ($m_{X_{b\bar{b}}}> m_\Upsilon$), its decay constant is one
third of that of $\Upsilon$: $f_{X_{b\bar b}}=\frac{1}{3}f_\Upsilon$
and its main decay mode is similar to that of the particle $X(4260)$
\cite{PDG}, i.e., $X_{b\bar{b}}\to \Upsilon \pi\pi$, but not $X_{b
\bar{b}}\to J/\psi \pi\pi$ so same as $X(4260)$. Finally we obtain
the total cross-section of producing $X_{b\bar{b}}$ at $Z_0$ peak is
$1.898\cdot 10^{-3} pb$; while as a comparison, the total
cross-cross section of producing $\Upsilon$ is $5.650\cdot 10^{-2}
pb$.

{\bf Summary:} From the results obtained here, one may see that the
production of a particle with Quantum numbers $J^{PC}=1^{--}$ via
the two-body final state processes $e^+ + e^- \rightarrow {\gamma} +
H_{Q\bar{Q}}$ concerned here is of ISR essentially and the total
cross sections for the production of the heavy quarkonium states at
$Z$-boson peak via the two-body final state processes can be so
great $\sigma\simeq 1\cdot 10^{-1}pb$ in magnitude, when
$H_{Q\bar{Q}}$ is a heavy quarkonium with quantum numbers
$J^{PC}=1^{--}$; so great $\sigma\simeq 1\cdot 10^{-2}pb$, when
$H_{Q\bar{Q}}$ is a `X(4260)-like' particle with quantum numbers
$J^{PC}=1^{--}$ as assumed in the above; whereas the total cross
sections at $Z$-boson peak for producing the quarkonium states
$H_{Q\bar{Q}}$ with the other quantum numbers (not $J^{PC}=1^{--}$)
via the two-body final state processes (without ISR) can only be
$\sigma\simeq 10^{-3}\sim 10^{-4}pb$ in magnitude. Therefore, one
can conclude that numerous events of $H_{Q\bar{Q}}$ with
$J^{PC}=1^{--}$ (the heavy quarkonia and even a `$X(4260)$-like
particle such as $X_{b\bar b}$ as well) may be produced at a
$Z$-factory. Also at such a $Z$-factory, hundreds, even thousands
events of the heavy quarkonia $H_{Q\bar{Q}}$ with different quantum
numbers from $J^{PC}=1^{--}$ may be produced. Considering the
advantage of the mono-energy characteristic for the produced photon
and heavy quarkonium both in the two-body final state processes, as
well as the clean environment in $e^+e^-$ collision, it seems that
it is worth more efforts quantitatively to estimate the precise
advantages in studying the production and the spectrum of the heavy
quarkonium systems additionally with specific condition of a
synthesis detectors at $Z$-factory. One can also conclude that it
will be better if the luminosity of the $Z$-factory may reach to so
high ${\cal L}\geq 10^{35\sim 36}cm^{-2}s^{-1}$ for the heavy
quarkonium studies. Moreover in Ref.\cite{cww}, we report additional
advantages in studying the heavy quarkonium properties at a
$Z$-factory, but via some processes else.

{\bf Acknowledgement:} This work was supported in part by Natural
Science Foundation of China (NSFC) under Grant No.10805082,
No.10875155, No.10847001, No.10875155 and No.10775141 and by Natural
Science Foundation Project of CQ CSTC under Grant No.2008BB0298.
This research was also supported in part by the Project of Knowledge
Innovation Program (PKIP) of Chinese Academy of Sciences, Grant No.
KJCX2.YW.W10.

\end{document}